%% file: article.tex
\documentclass[aps,prl,twocolumn,showpacs,preprintnumbers,superscriptaddress,nofootinbib,floatfix,10pt]{revtex4-2}

\usepackage{amsfonts,amssymb,stmaryrd,latexsym,amsmath,braket}
\usepackage{graphicx,subfigure}
\usepackage{comment}
\usepackage{times}
\usepackage{slashed}
\usepackage{bm}
\usepackage{color}
\usepackage{natbib}
\usepackage[colorlinks,linkcolor=blue,anchorcolor=red,urlcolor=blue,citecolor=red]{hyperref}

\usepackage{bibunits}

\newcommand{\beginsupplement}{%
        \setcounter{table}{0}
        \renewcommand{\thetable}{S\arabic{table}}%
        \setcounter{figure}{0}
        \renewcommand{\thefigure}{S\arabic{figure}}%
        \setcounter{equation}{0}
        \renewcommand{\theequation}{S\arabic{equation}}%
     }

\makeatletter
\let\saved@includegraphics\includegraphics
\AtBeginDocument{\let\includegraphics\saved@includegraphics}
\makeatother

\begin{document}

\begin{bibunit}[apsrev4-2]

\title{Lattice simulation of nucleon distribution and shell closure in the proton-rich nucleus $^{22}$Si}

\author{Shuang Zhang}
\affiliation{Institute for Advanced Simulation (IAS-4), Forschungszentrum J\"{u}lich, D-52425 J\"{u}lich, Germany}

\author{Serdar Elhatisari}
\affiliation{Interdisciplinary Research Center for Industrial Nuclear Energy (IRC-INE), King Fahd University of Petroleum and Minerals (KFUPM), 31261 Dhahran, Saudi Arabia}
\affiliation{Faculty of Natural Sciences and Engineering, Gaziantep Islam Science and Technology University, Gaziantep 27010, Turkey}
\affiliation{Helmholtz-Institut f\"{u}r Strahlen- und Kernphysik and Bethe Center for Theoretical Physics, Universit\"{a}t Bonn, D-53115 Bonn, Germany}

\author{Ulf-G. Mei{\ss}ner}
\affiliation{Helmholtz-Institut f\"{u}r Strahlen- und Kernphysik and Bethe Center for Theoretical Physics, Universit\"{a}t Bonn, D-53115 Bonn, Germany}
\affiliation{Institute for Advanced Simulation  (IAS-4), Forschungszentrum J\"{u}lich, D-52425 J\"{u}lich, Germany}
\affiliation{Peng Huanwu Collaborative Center for Research and Education, Beihang University, Beijing 100191, China}
\affiliation{Tbilisi State University, 0186 Tbilisi, Georgia}

\author{Shihang Shen}
\affiliation{Peng Huanwu Collaborative Center for Research and Education, Beihang University, Beijing 100191, China}
\affiliation{Institute for Advanced Simulation (IAS-4), Forschungszentrum J\"{u}lich, D-52425 J\"{u}lich, Germany}

\begin{abstract}

The proton-rich nucleus $^{22}$Si is studied using Nuclear Lattice Effective Field Theory with high-fidelity chiral forces. Our results indicate that $^{22}$Si is more tightly bound than $^{20}$Mg, thereby excluding the possibility of two-proton emission. 
The $Z = 14$ shell closure in $^{22}$Si is supported by the evolution of the $2^+$ state in the neighboring nuclei.
We then focus on the charge radius and spatial distribution information of $^{22}$Si, considering the novel phenomena that may emerge due to the small two-proton separation energy and the shell closure.
We present the distribution of the $14$ protons and $8$ neutrons obtained from our lattice simulation, revealing insights into the spatial arrangement of the nucleons. Moreover, the spatial localization of the outermost proton and neutron suggests that $^{22}$Si is a doubly magic nucleus.
Furthermore, we develop the pinhole method based on the harmonic oscillator basis, which gives insight into the nuclear structure in terms of the shell model picture from lattice simulations. Our calculated occupation numbers support that $Z = 14$ and $N = 8$ are the shell closures and show that the $\pi 1s_{1/2}$ orbital component is minor in $^{22}$Si.
\end{abstract}
\maketitle
\date{today}

\paragraph{}{\itshape Introduction.} Nuclei exhibiting a significant neutron-proton imbalance reveal novel insights into nuclear structure and nuclear astrophysics, but they pose major challenges to theoretical approaches. Recent discoveries of a growing number of proton-rich nuclei have significantly impacted our understanding of nuclear symmetries \cite{PhysRevLett.122.122501,PhysRevLett.129.122501,PhysRevLett.129.242502}, shell evolution
\cite{PhysRevLett.131.092501,PhysRevLett.127.262502}, astrophysical processes \cite{PhysRevLett.106.252503},
and proton emission \cite{PhysRevLett.131.172501,PFUTZNER2023104050}. Notably, the locations of the proton dripline for the P, S, and Ar elements have been accurately determined through B$\rho$-defined isochronous mass spectrometry experiments \cite{yu2024}. Moving further down the nuclear chart, the nucleus $^{22}$Si emerges as a particularly intriguing case due to its proximity to the proton dripline and its unique occupancy of the $Z = 14$ subshell closure and $N = 8$ shell closure.
It is probably the lightest bound nucleus with the $T_Z = -3$, where $T$ denotes the nuclear isospin. 
Nuclei located at shell closures feature enhanced stability compared to their neighbors and serve as cornerstones
for our understanding of nuclear structure within the nuclear landscape. 
It is also interesting to study the
evolution in exotic regions of the chart of nuclides under the combined actions of nuclear forces and many-body correlations. 
Hence, we focus on $^{22}$Si as the first test case for Nuclear Lattice Effective Field Theory (NLEFT) \cite{Lahde:2019npb} calculations of the proton-rich region, which can probe the isospin symmetries and the evolution of certain properties in
exotic regions of the nuclear chart, and provide essential tests to the nuclear forces and many-body correlations.

Much remains to be explored regarding the properties of $^{22}$Si, both from experimental and theoretical perspectives. It was initially measured at GANIL \cite{PhysRevLett.59.33}, and
its spectrum was subsequently studied \cite{PhysRevC.54.572}. Recent experiments have indirectly measured its ground-state (g.s.) mass
\cite{XU2017312,babo:tel-01461303}, but inconsistencies in the resulting two-proton emission properties necessitate further direct experimental measurements and theoretical studies. From the theoretical side,
{\it ab initio} calculations also suggest varying predictions for the two-proton separation energy
\cite{PhysRevLett.110.022502,PhysRevLett.126.022501,ZHANG2022136958}. Moreover, the large proton-neutron asymmetry ($\lvert N - Z
\rvert = 6$) in $^{22}$Si suggests that its charge radius could provide new constraints on the slope $L$ of the symmetry
energy in the equation of state \cite{PhysRevLett.127.182503,Konig:2023rwe} when combined with data for
its mirror partner $^{ 22}$O. 
In addition, the isospin asymmetry observed in the $^{22}$Si/$^{22}$O Gamow-Teller transition \cite{PhysRevLett.125.192503}
and the first mass measurement of $^{22}$Al \cite{Sun2024,PhysRevLett.132.152501}, respectively, revealed and suggested
the halo structure in $^{22}$Al. Given that $^{22}$Si lies nearer the edge of the nuclear landscape, it is fascinating
to explore how its shell closure and proton-neutron imbalance compete, and to investigate the  $\pi 1s_{1/2}$
component in $^{22}$Si, which also gained a lot of attention in studies of the bubble nuclei $^{34}$Si on the neutron-rich side of silicon isotopes \cite{Mutschler2017}.  

In this work, we perform Monte Carlo simulations for $^{22}$Si in the framework of NLEFT with high-fidelity
nucleon-nucleon (NN) and three-nucleon (3N) forces \cite{Elhatisari:2022zrb}. Using the pinhole algorithm \cite{Elhatisari:2017eno} and
the newly proposed pinhole algorithm based on the harmonic oscillator (HO) basis, we are able to probe the inner
structure of $^{22}$Si. Using these theoretical methods, we compute the binding energies and first $2^+$ states
of $^{22}$Si and its neighboring nuclei with a global chiral NN and 3NF interaction \cite{Elhatisari:2022zrb}, as well as an interaction specifically fine-tuned for silicon isotopes \cite{Konig:2023rwe}, both at next-to-next-to-next-to-leading order (N$^3$LO). We suggest that $^{22}$Si is a dripline nucleus and discuss
the $Z = 14$ shell using the evolution of the $2^+$ states in $N = 8$ isotones. We also give predictions
for the radii of $^{22}$Si, and show the charge radius difference with its mirror partner $^{22}$O. Finally, we
focus on the nucleon distributions of $^{22}$Si and present the occupation numbers for the g.s. of $^{22}$Si.

\paragraph{}{\itshape Methods.} NLEFT has the advantage of capturing  the full set of many-body correlations
with the use of auxiliary-field Monte Carlo simulations, but suffers from severe sign problems in its application
to high-fidelity chiral forces described by the chiral Hamiltonian $H_\chi$. In Ref. \cite{Elhatisari:2022zrb},
the wavefunction matching method was developed which maps the unitarily transformed $H_\chi' = UH_\chi U$
to a simple Hamiltonian $H_S$, where the $H_S$ is largely free of sign oscillations. This simple Hamiltonian
consists of smeared two-nucleon contact interactions as well as regularized one-pion exchange. If $H_\chi'$
is sufficiently close to  $H_S$, first-order perturbation theory can be used efficiently to calculate the
higher-order chiral forces up-to-and-including N$^3$LO. Furthermore, smeared 3NFs are then fitted to masses of selected nuclei ranging from $A=3$ to $A=40$. Consequently, nuclear charge radii as well the equation of state
for neutron and for nuclear matter can be predicted,
showing good agreement with the data. We use the same Hamiltonian in this study and refer to Ref.~\cite{Elhatisari:2022zrb}
for further details. To further study the systematical uncertainties of our calculations, we also use the modified version of the 3NFs that was fine-tuned to the binding energies of the silicon isotopic chain \cite{Konig:2023rwe}, and we refer to the chiral force from Ref.~\cite{Elhatisari:2022zrb} as global interaction in what follows.

In this work, we perform simulations on a periodic cubic lattice with a side length of $L = 10$ and lattice spacing $a \simeq 1.32\,$fm, which
corresponds to a momentum cutoff $\Lambda = \pi/a \simeq 471$~MeV. The temporal lattice spacing is taken to be $a_t
= 0.001$~MeV$^{-1}$ to ensure the convergence of the calculation. We perform local smearing with the parameter
$s_{\text{L}} = 0.07$, the nonlocal smearing $s_{\text{NL}} = 0.5$, and coupling constant $c_{{\rm SU(4)}} = -4.6\times 10^{-7}$~MeV$^{-2}$
for the NN contact interaction in $H_S$. 
We use the evolution operator in Euclidean time, $\exp(-H_{S}\tau)$ to extract the states of
interest. The initial states are chosen as products of single-particle HO wavefunctions. Subsequently, the chiral Hamiltonian $H_\chi'$ is calculated pertubatively \cite{LU2019134863}.  All the calculated energies and radii are then extrapolated in Euclidean time using the formalism from
Refs.~\cite{Lahde_2015,PhysRevA.101.063615,Shen2023}, while the nucleon distribution and occupation numbers are
simulation results at $\tau = 0.2$~MeV$^{-1}$.

\begin{figure}[t]
\centering
  \includegraphics[width=0.48\textwidth]{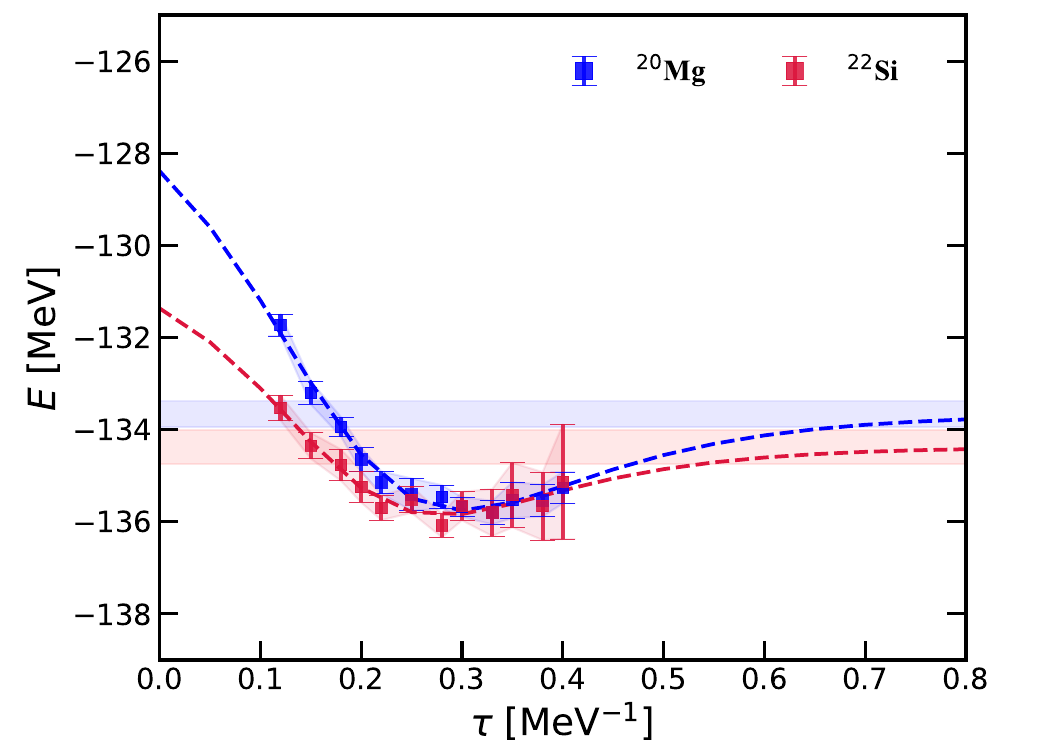}\\
  \caption{Calculated g.s. energies of $^{22}$Si and $^{20}$Mg using the chiral forces at N$^3$LO \cite{Elhatisari:2022zrb}. The red and blue squares represent the calculated results of $^{22}$Si
    and $^{20}$Mg varying with Euclidean time $L_t$, respectively, while the experimental value for
    $^{20}$Mg is shown as the black dash-dotted line \cite{ensdf}. The error bars and shaded regions indicate the statistical uncertainties in the calculations. The red and blue dashed lines correspond to the extrapolated curves for $^{22}$Si
    and $^{20}$Mg. The extrapolated results are presented with error bands centered on the corresponding extrapolated energies.}
  \label{fig:BE}
\end{figure}

For the radii and nucleon distribution calculations, we employ the pinhole algorithm \cite{Elhatisari:2017eno},
details are presented in~\cite{supp}. 
To gain more insight into the nuclear structure, we propose here a variation of the pinhole method based on the
HO basis to estimate the occupation numbers of shell model orbitals.
With this approach, we aim to establish a bridge between the Monte Carlo simulations in coordinate or
momentum space and the shell model in configuration space.
Instead of sampling the nucleon positions and spin-isospin as proposed in the standard pinhole algorithm, we sample the HO basis characterized by the principal, orbital angular momentum, total angular momentum, and angular momentum projection quantum number.

\paragraph{\itshape Results.} 
In Fig.~\ref{fig:BE}, we present the calculated g.s. energies of $^{22}$Si and $^{20}$Mg with different $\tau$ and extrapolate to  infinite Euclidean time. 
We get $-133.66 (28)$~MeV and $-134.38 (39)$~MeV for $^{20}$Mg and $^{22}$Si, respectively, 
whereas experiment gives $-134.61$~MeV for $^{20}$Mg~\cite{PhysRevLett.113.082501}.
For $^{22}$Si, there is no direct experimental measurement, however, the indirect measurement gives $-134.51$~MeV \cite{XU2017312}.
The $\beta$-delayed one-proton and two-proton emissions in $^{22}$Si have been measured, while two-proton decay has
not yet been ruled out \cite{XU2017312,babo:tel-01461303}. The location of the dripline in the mirror $Z = 8$ oxygen
isotopes provides validation for several many-body methods and highlights the significance of three-body forces in
chiral EFT \cite{PhysRevLett.105.032501,PhysRevLett.108.242501,PhysRevLett.110.242501}. Similarly, the dripline of
proton-rich $N=8$ isotones presents an intriguing area for further investigation. Indeed, many-body calculations
based on chiral forces yield varying predictions for the two-proton separation energies of $^{22}$Si
\cite{PhysRevLett.110.022502,PhysRevLett.126.022501,ZHANG2022136958}.
Our calculations indicate that $^{22}$Si is more bound than $^{20}$Mg within the statistical error. Specifically,
when the maximum $\tau$ is increased from $0.3$ MeV$^{-1}$ to $0.4$ MeV$^{-1}$, the extrapolated g.s. energies
for both nuclei change by less than $0.36$~MeV, and $^{22}$Si consistently remains more bound than $^{20}$Mg.
It yields a two-proton separation energy of $S_{2p} = 0.72 (48)$~MeV with maximum $\tau = 0.4$~MeV$^{-1}$,
which suggests that $^{22}$Si is not a $2p$ emission candidate. Furthermore, we also employed an alternative set of chiral forces proposed for silicon isotopes for a sensitivity analysis \cite{Konig:2023rwe}.
The calculated g.s. energies of $^{22}$Si and $^{20}$Mg are $-136.28 (41)$~MeV and $-134.94 (18)$~MeV, respectively.
The extracted $S_{2p}$ is $1.34 (45)$~MeV, which is consistent with the prediction from the global interaction.

\begin{figure}[t]
\centering
  \includegraphics[width=0.48\textwidth]{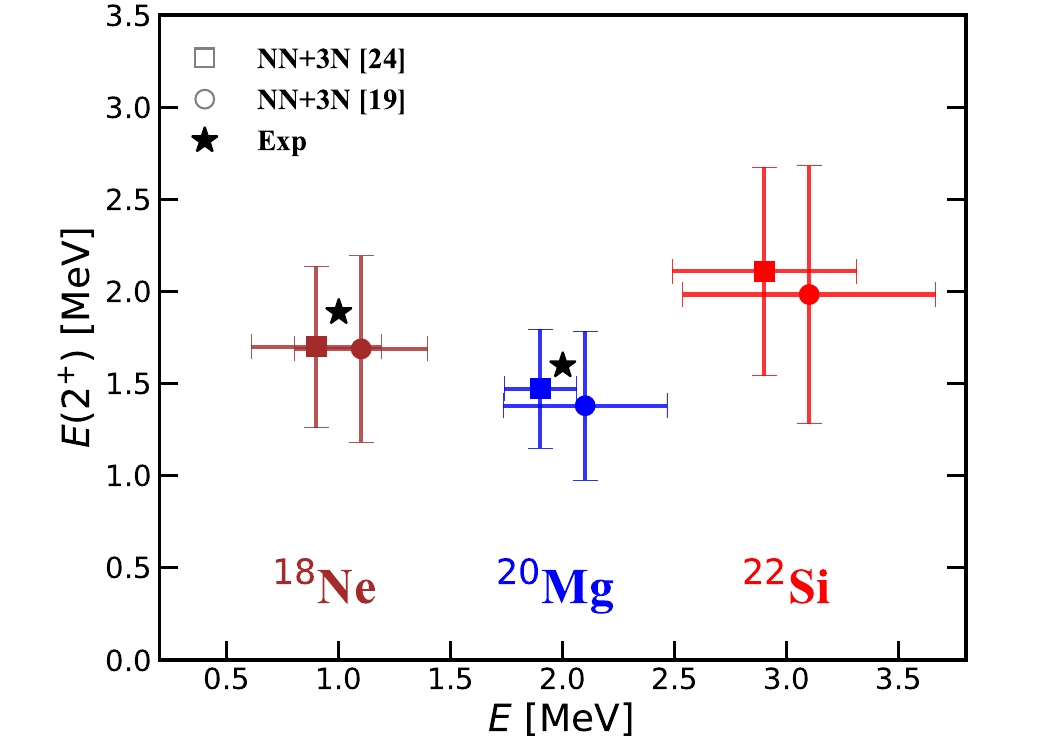}\\
  \caption{Excitation energies of $J = 2^{+}$ states in the $N = 8$ isotones $^{18}$Ne, $^{20}$Mg and $^{22}$Si
    (experiment \cite{ensdf}: black stars, theory: squares and circles). The results based on the global interactions
    are shown by squares, whereas the results based on interactions for silicon isotopes are presented by circles.
    The vertical solid lines refer to the uncertainties in the excitation energies of the $2^+$ states,
    while the horizontal solid lines indicate the uncertainties in the absolute energies of the $2^+$ states due to the Euclidean time extrapolation. Note that these squares,
    circles, and stars are plotted as references on the $E$-axis, and only the magnitude of the horizontal
    errors are calibrated by the values on the $E$-axis.}
  \label{fig:shell}
\end{figure}

Due to the fact that $^{22}$Si is located at the edge of the nuclear landscape and its mirror partner, $^{22}$O,
is a well-known doubly magic nucleus, we evaluate the potential of shell closure in $^{22}$Si.
We then focus on the evolution of $2^+$ states in $^{22}$Si and its nearby $N = 8$ isotones as a key indicator
of the $Z = 14$ shell closure. In Fig.~\ref{fig:shell}, we show the calculated $2^+$ states of $^{18}$Ne, $^{20}$Mg and
$^{22}$Si with two sets of chiral forces, and compare with experiment \cite{ensdf}. The energies of the $2^+$ states in $^{18}$Ne,
$^{20}$Mg, and $^{22}$Si based on the global interaction are  $1.70 (44)$~MeV, $1.47 (32)$~MeV, and $2.11 (57)$~MeV,
respectively, while the results based on the alternative interaction are $1.69 (51)$~MeV, $1.38 (40)$~MeV,
and $1.98 (70)$~MeV, respectively. The energies of  $2{^+}$ states extracted from these two calculations
are in good agreement. While a measurement for $2{^+}$ state in $^{22}$Si is not yet available, the calculated $2{^+}$
states of $^{18}$Ne and $^{20}$Mg agree with the experimental values of $1.89$~MeV and $1.60$~MeV. The energies
of the $2^+$ states in $^{22}$Si 
are higher than for the other $N = 8$ isotones, suggesting a possible $Z = 14$ subshell closure at
the proton dripline. Since $N = 8$ is a conventional magic number, proton excitation may challenge the notion of estimating the neutron shell structure in $Z = 14$ isotopes through the energies of the $2^+$ state \cite{Li:2023bel}. We will later use the pinhole method to calculate the nucleon distributions and occupation numbers on the HO basis. Meanwhile, these results provide valuable insights into the characteristics of the $N = 8$ shell.

\begin{figure*}[t]
    \centering
    \begin{minipage}[b]{0.495\textwidth}
        \centering
            \includegraphics[width=\textwidth]{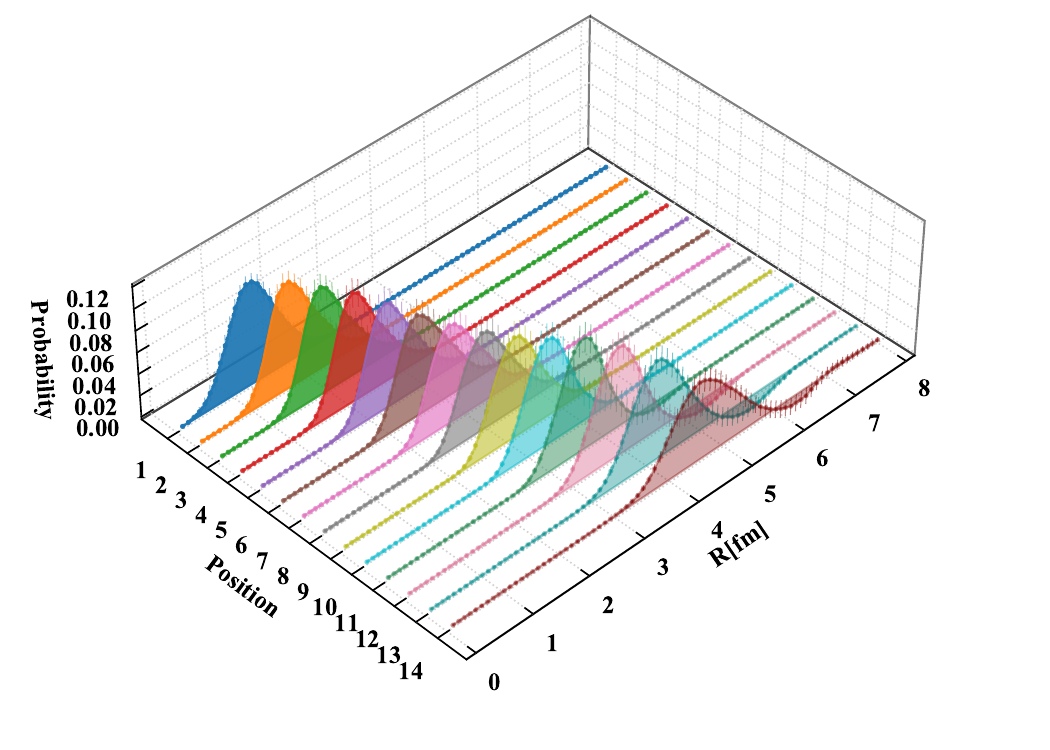}
    \end{minipage}
    \hspace{0.0001\textwidth}
    \begin{minipage}[b]{0.495\textwidth}
        \centering
        \includegraphics[width=\textwidth]{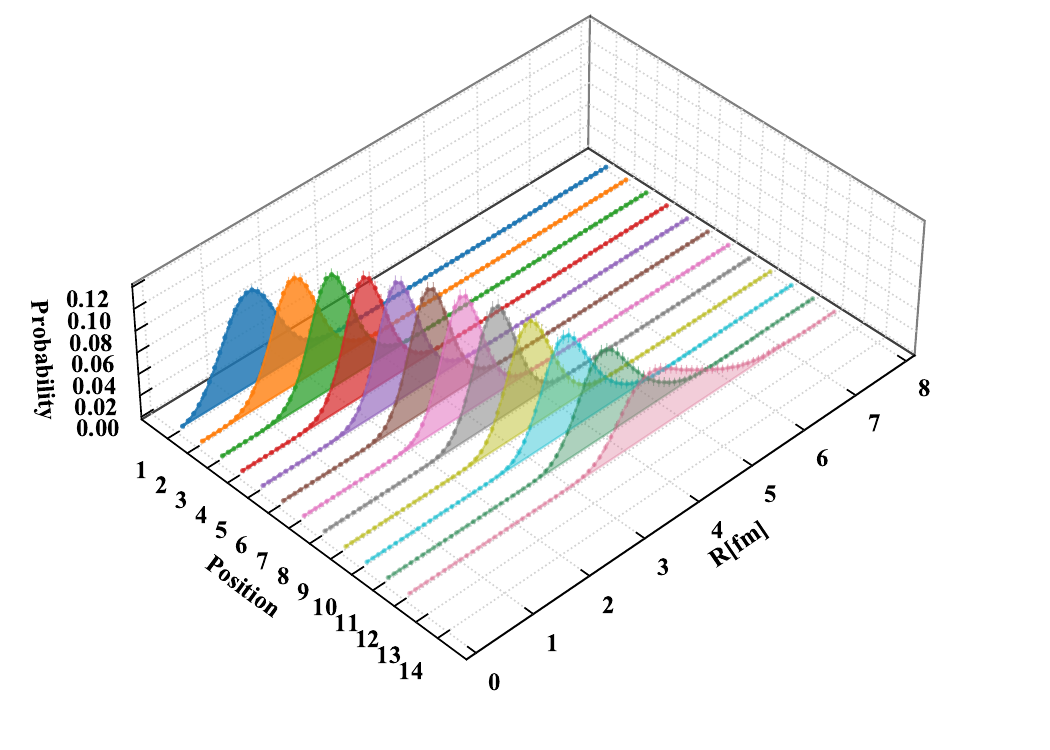}
    \end{minipage}
    \caption{Proton distribution probabilities in $^{22}$Si (left panel) and $^{20}$Mg (right panel) simulated using the pinhole method with the global chiral
    force at N$^3$LO. In each simulation,
    the distance of $Z$ protons from the CoM is arranged in ascending order. The colors represent the distribution of $Z$ groups of protons, ranging from near to far relative to the CoM. The uncertainty in the simulation is represented by the error bar.}
    \label{fig:Dis}
\end{figure*}

Recent experiments reveal the halo structure in the excited state of $^{22}$Al \cite{PhysRevLett.125.192503}, and
suggest that the $^{22}$Al is a proton dripline halo candidate \cite{PhysRevLett.132.152501}. As the $T = 3$
isobaric analogue pair of $^{22}$Al, $^{22}$Si is closer to the edge of the nuclear landscape and fully occupies
the shell. The interplay between proton-neutron imbalance and shell effects makes the values of the radii
in this exotic region particularly intriguing to study. Consequently, we use the pinhole method to calculate
the radius of the $^{22}$Si g.s. and its radius difference with neighboring nuclei. The charge radius
is computed using the formula
$R^2_{\mathrm{ch}} = \langle r^2_{\mathrm{pp}}  \rangle + R^2_{\mathrm{p}} + (N/Z) R^2_{\mathrm{n}} + r^2_{\mathrm{DF}}$,
where $\langle r^2_{\mathrm{pp}}  \rangle$ is the mean-square point proton radius, intrinsic proton and neutron radius
$R^2_{\mathrm{p}}$ and $R^2_{\mathrm{n}}$ is $0.7056$~fm$^2$ \cite{PhysRevLett.128.052002} and $-0.105$~fm$^2$
\cite{PhysRevC.103.024313}, respectively, and the Darwin-Foldy term $r^2_{\mathrm{DF}}$ is $0.033$~fm$^2$.
In Tab.~\ref{tab:radii}, we present the calculated charge radii $R_{\text{ch}}$ for the $^{22}$Si g.s.
from the global chiral forces \cite{Elhatisari:2022zrb} and the chiral forces fine-tuned for
the silicon isotopes~\cite{Konig:2023rwe}. There are slight differences between these results. We combine them so that we obtain an estimate of the radius at the $3\sigma$ confidence level, ranging from 3.145 to 3.334~fm. To illustrate the physical significance of these calculated radii, we further extract the charge radius differences $ \Delta R_{\text{ch}}$ between $^{22}$Si and $^{20}$Mg. The calculated radii of $^{20}$Mg with
these interactions are $3.207 (24)$~fm, and $3.126 (22)$~fm, respectively. We find the
$ \Delta R_{\text{ch}}$ between $^{22}$Si and $^{20}$Mg obtained by two sets of chiral forces are consistent.
The extracted $\Delta R_{\text{ch}}$ between $^{22}$Si and $^{20}$Mg is 0.070~fm, which is comparable to the $\Delta R_{\text{ch}} ({^{28}\text{Si}\text{-}{^{26}\text{Mg}}}) = 0.088$~fm,
$\Delta R_{\text{ch}}  ({^{30}\text{Si}\text{-}{^{28}\text{Mg}}}) = 0.064$~fm and
$\Delta R_{\text{ch}}  ({^{32}\text{Si}\text{-}{^{30}\text{Mg}}}) = 0.042$~fm measured in  experiments
\cite{PhysRevLett.108.042504,Konig:2023rwe}. 
The $\Delta R_{\text{ch}}$ between these isotopes of silicon and magnesium is of a similar magnitude, indicating
similar underlying nuclear structure effects may influence the charge distribution in
\begin{table}[ht]
\centering
\caption{Charge radii ($ R_{\text{ch}}$), charge radius difference $ \Delta R_{\text{ch}} ({^{22}\text{Si}\text{-}{^{20}\text{Mg}}})
  = R_{\text{ch}}(^{22}\text{Si}) - R_{\text{ch}}({^{20}\text{Mg}})$ and mirror charge radius difference $ \Delta R_{\text{ch}}
  ({^{22}\text{Si}\text{-} {^{22}\text{O}}}) = R_{\text{ch}}(^{22}\text{Si}) - R_{\text{ch}}(^{22}\text{O})$ of $^{22}$Si calculated
  with the  global chiral forces \cite{Elhatisari:2022zrb} and chiral  forces fine-tuned
  for the silicon isotopes~\cite{Konig:2023rwe}. All values are given in fm and include statistical errors.}
\begin{tabular}{cccc}
\hline
\hline
& $ R_{\text{ch}}$ & $\Delta R_{\text{ch}} ({^{22}\text{Si}\text{-}{^{20}\text{Mg}}})$ & $\Delta R_{\text{ch}}
({^{22}\text{Si}\text{-}{^{22}\text{O}}})$ \\
 \hline
$H_{\chi}$ \cite{Elhatisari:2022zrb} & 3.277 (19) & 0.070 (31) & 0.361 (32) \\ \hline
$H_{\chi}$ \cite{Konig:2023rwe} & 3.196 (17) & 0.070 (28) & 0.370 (18) \\ \hline\hline
\end{tabular}
\label{tab:radii}
\end{table}
the dripline nucleus $^{22}$Si, despite its relatively small $S_{2p}$. 
In addition, we calculated the mirror charge radius differences for the $^{22}\text{Si}$-${^{22}\text{O}}$ mirror pair.
The correlation between mirror charge radius difference and the slope $L$ in the symmetry energy was found to
be $\Delta R_{\text{ch}} \propto \lvert N - Z \rvert \times L$ in the density function theory calculation
\cite{PhysRevLett.119.122502}, and has been used to constrain the $L$
\cite{PhysRevResearch.2.022035,PhysRevLett.127.182503}. With a large proton-neutron imbalance $\lvert N - Z
\rvert = 6$, the $\Delta R_{\text{ch}}$ between the $^{22}\text{Si}$-${^{22}\text{O}}$ mirror pair will significantly
constrain $L$. 
The calculated mirror charge radii
between the $^{22}\text{Si}$-${^{22}\text{O}}$ mirror pair with chiral NN+3N forces show good agreement. 
Coupled cluster and the auxiliary field
diffusion Monte Carlo method in continuum space have established a linear regression relationship between
the charge radius difference $\Delta R_{\text{ch}}$, and the isospin difference based on the computational results
\cite{PhysRevLett.130.032501}. For the $^{22}\text{Si}$-${^{22}\text{O}}$ mirror pair, this linear regression
yields a charge radius difference of $\Delta R_{\text{ch}} = 0.429 (21)$~fm. Our lattice simulation results are
relatively small but remain consistent with this prediction within the $3\sigma$ confidence level. Here we also give the slope
$L = 56.4(5.8)$~MeV extracted based on the global interaction \cite{Elhatisari:2022zrb} for reference.

We use the pinhole method \cite{Elhatisari:2017eno} to simulate the position distribution of $A$ nucleons. In order to reduce lattice artifacts, each pinhole configuration was smoothed using $10^3$ iterations of a Gaussian smearing function parameterized by the proton size, resulting in a more realistic and continuous nucleon distribution.
In each simulation, the nucleon coordinates are rearranged from small to large based on the distance of each nucleon relative to the center-of-mass (CoM).
In Fig.~\ref{fig:Dis}, we show the simulated distribution of the $14$ protons in $^{22}$Si and the $12$ protons
in $^{20}$Mg based on the  global chiral force at  N$^3$LO \cite{Elhatisari:2022zrb}. 
Due to the sign problem, the distributions of the outermost nucleons sometimes become negative, however,
within the bounds of the statistical uncertainty, and these values are very close to zero.

Since we use Gaussian smearing, the overall nucleon distribution tends to be Gaussian, but we find that the outermost proton in $^{20}$Mg tends to be more extended than the other protons.
As shown in~\cite{supp}, we also observe this phenomenon in the outermost neutron distribution of $^5$He and $^6$He.
The degree of nucleon binding is reflected in the extension of the outermost nucleon distribution. 
In the simulation of $^{22}$Si using the high-fidelity forces, we observe that the outermost proton distribution is similar to that of the inner layers, resulting in a less extended spatial distribution compared to that observed in  $^{20}$Mg. 
Moreover, the outermost neutron distributions in $^{22}$Si exhibit spatial localization characteristics and show a similar broadening pattern to those of the inner nucleon distributions, as also shown in~\cite{supp}.
These spatial localizations reveal, to a certain extent, the presence of shell closure at $Z = 14$ and $N = 8$
in $^{22}$Si, indicating that $^{22}$Si is a doubly magic nucleus at the proton dripline.
Additionally, we find in our calculations that the
general features of the chiral force significantly enhance the $Z = 14$ shell closure compared to the simple Hamiltonian. As detailed in~\cite{supp}, we perform an in-depth analysis of each set of proton distributions by calculating the average distances derived from these distributions. This study reveals the structural similarity between the proton distributions of $N = 8$ isotones $^{22}$Si and $^{20}$Mg, explores the preferred spatial positions of the two extra protons in $^{22}$Si, and demonstrates that the outmost proton in $^{22}$Si is more tightly packed in space.

Using the pinhole method within the HO basis, we calculated the HO orbitals occupation numbers of $^{22}$Si with the empirical parameter $\hbar\omega = 41A^{-1/3}$ MeV and interactions based on the global chiral nuclear force. These calculations provide a detailed insight into the nucleon distributions across the HO orbitals, revealing the underlying shell structure of $^{22}$Si. In~\cite{supp}, we present several HO orbital occupation numbers of neutrons ($n_{\nu}$) and protons ($n_{\pi}$) in $^{22}$Si. For the neutron occupation numbers, it is evident that the $N = 8$ shell is fully occupied, indicating a minimal contribution from higher HO orbitals and confirming the presence of the $N = 8$ shell closure. Regarding the proton occupation numbers, the $Z = 14$ shell is also nearly fully occupied. The six valence protons predominantly occupy the $0d_{5/2}$ orbital, though with relatively larger statistical errors compared to the eight protons in the inner shell. These results further support our conclusions regarding the existence of shells from the spatial distribution of nucleons. Furthermore, the low occupancy of the $1s_{1/2}$ orbital reflects a minor proton $1s$-wave component on the proton-rich side of the silicon isotopes. This also indicates the closed-shell effect in $^{22}$Si remains dominant, outweighing the impact of the proton-neutron imbalance at the landscape boundary.

\paragraph{}{\itshape Summary.} We have used  NLEFT to understand the structure of the $^{22}$Si nucleus. Using
two sets of chiral NN and 3N forces at N$^3$LO, we calculated the binding energies and $2^{+}$
excitation energies of $^{22}$Si, $^{20}$Mg, and $^{18}$Ne.
Overall results are in agreement with the available experimental data and
predict that $^{22}$Si is a dripline nucleus, as well as confirm the presence of a shell closure at $Z = 14$ in
$^{22}$Si. Additionally, we calculated the charge radius of $^{22}$Si, and extracted the mirror charge radius
difference, which can serve as a reference for future experiments. 
It was also found that the calculated radii of $^{22}$Si differ from $^{20}$Mg in a manner similar to the
difference observed between silicon and magnesium isotopes in the proton-neutron balance region.
Finally, insights into the $^{22}$Si structure have been demonstrated in both coordinate and
configuration space using the pinhole method. 
We simulated the spatial distribution characteristics of nucleons and extracted the HO basis occupation
numbers in $^{22}$Si. It further supports the $^{22}$Si is a doubly magic nucleus and unveils detailed
nucleon information within the nucleus while accounting for the full set of many-body correlations.
While the work presented here offers a new perspective on exploring nuclear structure, many other ways
remain to be explored for a more comprehensive understanding.


\begin{acknowledgments}
We are grateful for discussions with Dean Lee and other members of the NLEFT collaboration.
This work is part of the EXOTIC grant and was supported in part by the European Research Council (ERC) under the European Union's Horizon 2020 research and innovation programme (grant agreement No. 101018170).
The work of UGM was also supported  by the CAS President's International
Fellowship Initiative (PIFI) (Grant No.~2025PD0022).
The work of SE is supported in part by the Scientific and Technological Research Council of Turkey (TUBITAK project no. 120F341).
The authors gratefully acknowledge the Gauss Centre for Supercomputing e.V. (www.gauss-centre.eu)
for funding this project by providing computing time on the GCS Supercomputer JUWELS
at J\"ulich Supercomputing Centre (JSC).
\end{acknowledgments}

\end{bibunit}
\clearpage

\beginsupplement

\input{supplemental_material}

\end{document}

%% file: supplemental_material.tex
\onecolumngrid
\begin{bibunit}[apsrev4-2]

\section*{Supplemental Material}
Nuclear Lattice Effective Theory (NLEFT) simulation costs scale modestly with the nucleon number $A$, $t_{\rm CPU} \sim
A^{1...2}$, due to the advantages of Monte Carlo sampling and the decoupling of the complicated many-body motion
into interactions between nucleons and auxliary background fields. 
This approach  captures the full set of many-body correlations, but it does not explicitly
confine the motion of the center-of-mass (CoM). Therefore information about density correlations relative to the CoM is lost.   
The pinhole method for determining $A$-body density correlations with respect to the CoM was
therefore developed in Ref.~\cite{Elhatisari:2017eno}. 
With the pinhole method, the exact locations of the $A$ nucleons can be determined by stochastically
sampling the nucleon positions, spins, and isospins for each pinhole configuration. Then the $A$-body spatial correlations
information relative to the CoM can be naturally obtained. Combining density correlations with any
observable of physical interest, we are able to probe the underlying structure of the nuclei.

\subsection{The pinhole method in coordinate space}
To describe the pinhole method, we first introduce the normal-ordered $A$-body density operator $\rho_{i_1,j_1,\cdots,i_A,j_A}(\mathbf{n}_1,\cdots,\mathbf{n}_A) = :\rho_{i_1,j_1}(\mathbf{n}_1)\cdots\rho_{i_A,j_A}(\mathbf{n}_A):$, where $\rho_{i,j}(\mathbf{n})$ is the one-body density operator for nucleons with spin $i$ and isospin $j$ at lattice site $\mathbf{n}$ in coordinate space. A key feature of the pinhole method is that the exact locations of nucleons with spin $i$ and isospin $j$ can be sufficiently sampled by inserting the normal-ordered $A$-body density operator.
Then, we insert the operator $\rho_{i_1,j_1,\cdots,i_A,j_A}(\mathbf{n}_1,\cdots,\mathbf{n}_A)$ at middle time of the Euclidean time projection amplitudes,
\begin{equation}
\bra{{\Psi_f}} M^{L_t/2} \rho_{i_1,j_1,\cdots, i_A,j_A} (\mathbf{n}_1, \cdots, \mathbf{n}_A) M^{L_t/2} \ket{\Psi_i},
\label{eq:smeq1}
\end{equation}
where $\ket{\Psi_i}, \ket{\Psi_f}$ are the initial and final trial wave functions, propagated via the transfer matrix $M =:\exp(-H_{S} a_{t}):$. The transfer matrix is connected with the evolution operator $\exp(-H_{S} \tau)$ via the Trotterization, which approximates the time evolution by dividing it into discrete steps for computational feasibility. Here, the difference $H_{\chi}'-H_{S}$ is calculated by perturbation theory, see Ref. \cite{LU2019134863} for more details. In the larger $\tau$ limit, we can get the physical states we want. 
Due to the completeness relation of the $A$-body density operator $\rho_{i_1,j_1,\cdots,i_A,j_A}(\mathbf{n}_1,\cdots,\mathbf{n}_A)$, the sum of this expectation value over $\mathbf{n}_1,\cdots,\mathbf{n}_A$ and $i_1,j_1,\cdots,i_A,j_A$ yields $A!$ times the amplitude $\bra{\Psi_f} M^{L_t}\ket{\Psi_i}$, which can be calculated using Auxiliary Fields Monte Carlo simulations.
In the auxiliary field formalism, the transfer matrix $M$ at each time slice depends on the auxiliary fields $s_{n_{t}}$, and the Eq. \eqref{eq:smeq1} can be rewritten in terms of the auxiliary fields. To evaluate it, we perform importance sampling according to the amplitude,
\begin{equation}
A(s;\mathbf{n}_1, \cdots, \mathbf{n}_A; i_1,j_1,\cdots, i_A,j_A) = |\bra{{\Psi_f}} M(s_{Lt})\cdots M(s_{Lt/2+1}) \rho_{i_1,j_1,\cdots, i_A,j_A} (\mathbf{n}_1, \cdots, \mathbf{n}_A)  M(s_{Lt/2})\cdots M(s_{1}) \ket{\Psi_i}|,
\end{equation}
which reflects the $A$-body density information with the auxiliary fields $s_{1},\cdots,s_{Lt}$, pinhole locations $\mathbf{n}_1,\cdots,\mathbf{n}_A$ and spin-isospins indices $i_1,j_1,\cdots,i_A,j_A$. To generate different amplitudes, the auxiliary fields $s_{1},\cdots,s_{Lt}$ are updated by the shuttle algorithm \cite{LU2019134863}, and then the pinhole locations $\mathbf{n}_1,\cdots,\mathbf{n}_A$ and spin-isospins indices $i_1,j_1,\cdots,i_A,j_A$ are sampled by the  Metropolis algorithm.
For any operator of interest, we can evaluate it by 

\begin{equation}
\langle\hat{O}\rangle =\frac{\sum{A(s;\mathbf{n}_1, \cdots, \mathbf{n}_A; i_1,j_1,\cdots, i_A,j_A)} \mathrm{exp}(i\theta[s;\mathbf{n}_1, \cdots, \mathbf{n}_A; i_1,j_1,\cdots, i_A,j_A])O{(\mathbf{n}_1, \cdots, \mathbf{n}_A; i_1,j_1,\cdots, i_A,j_A)}}{\sum{ A(s;\mathbf{n}_1, \cdots, \mathbf{n}_A; i_1,j_1,\cdots, i_A,j_A)}\mathrm{exp}(i\theta[s;\mathbf{n}_1, \cdots, \mathbf{n}_A; i_1,j_1,\cdots, i_A,j_A])},
\end{equation}
where the summation symbol indicates the sum of different auxiliary fields $s$, pinhole locations $\mathbf{n}_1,\cdots,\mathbf{n}_A$ and spin-isospins indices $i_1,j_1,\cdots,i_A,j_A$, and $\mathrm{exp}(i\theta[s;\mathbf{n}_1, \cdots, \mathbf{n}_A; i_1,j_1,\cdots, i_A,j_A])$ is the complex phase of each pinhole amplitude. 
In this paper, we focus on the spatial distribution of nucleons and, therefore, introduce the nucleon distance sorting operator,
\begin{equation}
\hat{\Pi}_{i}(r) =\delta(r-r_{\hat{\pi}(i)}),
\label{eq:smeq2}
\end{equation}
where $r$ denotes the radial distance of nucleon relative to the CoM, $i$ indicates the $i$-th closest nucleon to the CoM, and $\hat{\pi}$ is the position index operator, expressed as
\begin{equation}
\hat{\pi}(i) =\arg \min_{\substack{k \notin \{\pi(1),\ldots, \pi(i-1)\}}}\{\hat{r}_{k}\}.
\end{equation}
Here, $\hat{r}_k$ denotes the radial distance of the nucleon with index $k$ relative to the CoM in each pinhole configuration, $\min_{\substack{k \notin \{\pi(1),\ldots, \pi(i-1)\}}}$ identifies the minimum radial distance among those not already assigned in the previous $i-1$ steps, and $\arg$ return the index $k$ corresponding the minimum radial distance. 
The nucleon distance sorting operator arranges the distances of $A$ nucleons to the CoM from small to large.
Then, by measuring the expectation value of Eq. \eqref{eq:smeq2} with various pinhole configurations, we can determine the distribution of the $A$ group nucleons from nearest to farthest relative to the CoM.

\begin{figure*}[t]
\centering
  \includegraphics[width=\textwidth]{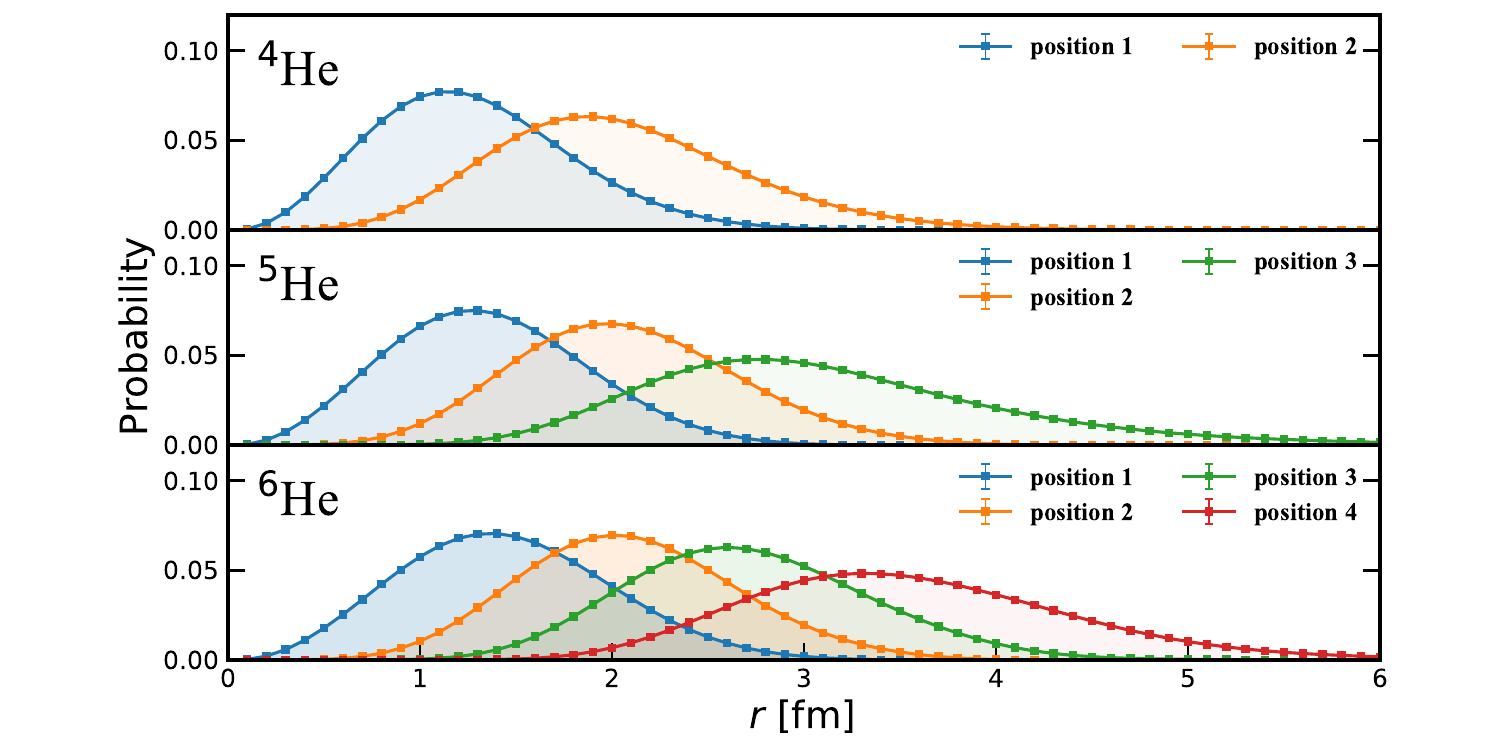}\\
  \caption{Calculated neutron distribution probabilities of $^{4,5,6}$He using the pinhole method with global chiral force at N$^3$LO.}
  \label{fig:He_dis}
\end{figure*}

We then verified our operator ansatz in the neutron distribution of $^{4-6}$He with chiral force at N$^{3}$LO, as shown in 
Fig.~\ref{fig:He_dis}. 
$^{4}$He is a stable nucleus, while the ground states of $^{5}$He and $^{6}$He are a resonance and exhibit halo phenomena, respectively. The distribution of the second-layer neutrons in $^{5}$He and $^{6}$He is remarkably close to that in $^{4}$He, and we can see that the outermost neutron distribution and the innermost neutron distribution of $^{4}$He have identical Gaussian distribution widths. These support the shell closure at $N = 2$. 
Furthermore, in $^{5}$He and $^{6}$He, due to the presence of unbound and weakly bound neutrons, the outermost neutron distribution has a significant spatial extension, which is markedly different from the distribution of the inner two layers of neutrons.
The observation in He isotopes demonstrates that our nucleon spatial distribution operator not only effectively reveals the spatial positions of nucleons but also reflects the shell closure to some extent.

\begin{figure*}[t]
\centering
  \includegraphics[width=0.6\textwidth]{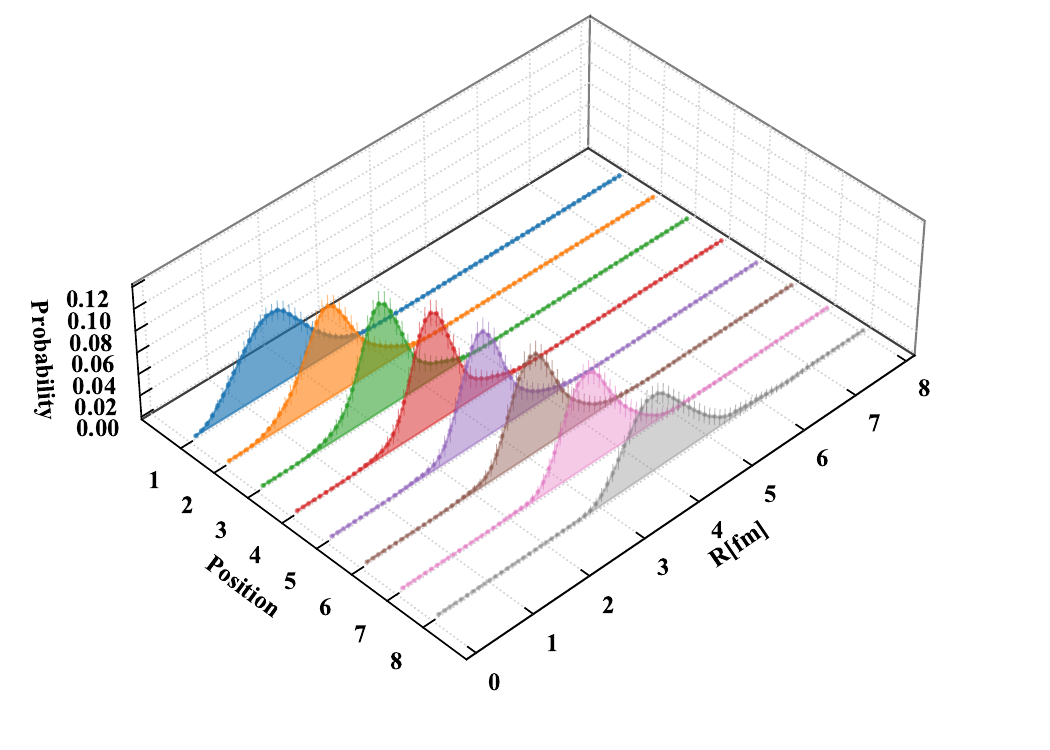}\\
  \caption{Calculated neutron distribution probabilities of $^{22}$Si using the pinhole method with global chiral force at N$^3$LO, similar to Fig.~\ref{fig:Dis}. }
      \label{fig:Si_ndis}
\end{figure*}

\subsection{The pinhole method in harmonic oscillator basis}

The harmonic oscillator (HO) basis is widely used in finite nuclei calculations since it is an angular-momentum eigenstate and can effectively describe the self-bound nature of atomic nuclei. We propose a pinhole method based on the HO basis that provides insights into nuclear structure in the shell model picture from NLEFT with Auxiliary Fields Monte Carlo simulations. In contrast to the standard pinhole method, we introduce a $A$-body HO basis in m-scheme at the middle time, 
\begin{equation}
\bra{{\Psi_f}} M^{L_t/2} \ket{\alpha_{1},\cdots,\alpha_{A}}\bra{\alpha_{1},\cdots,\alpha_{A}} M^{L_t/2} \ket{\Psi_i},
\label{eq:smhoeq1}
\end{equation}
where $\ket{\alpha}$ denotes the index of the single-particle HO basis. The HO basis space is truncated by $N_{\mathrm{shell}} =2n+l$, where $n$ is the principal quantum number and $l$ is the orbital angular momentum.
These basis states collectively form the many-body basis with good total angular momentum projections $M = \sum_i{m_i}$ and parity. To target a specific total angular momentum $J$, we select $M = J$. This ensures the basis contains configurations capable of representing the desired $J$ state.
As in the coordinate space pinhole method, here we calculate the Eq. \eqref{eq:smhoeq1} in auxiliary field formalism by performing importance sampling according to the occupation configurations of HO basis. For the generation of different auxiliary fields and occupation configurations, we update the auxiliary fields by the shuttle algorithm \cite{LU2019134863}, and then the occupation configurations of HO basis are sampled by the Metropolis algorithm. The occupation number in the HO basis can be calculated by
\begin{equation}
\langle\hat{N}(\alpha)\rangle =\frac{\sum{A(s;\alpha_{1},\cdots,\alpha_{A})} \mathrm{exp}(i\theta[s;\alpha_{1},\cdots,\alpha_{A}])\sum_{i}\delta_{\alpha, \alpha_i}}{\sum{A(s;\alpha_{1},\cdots,\alpha_{A})}\mathrm{exp}(i\theta[s;\alpha_{1},\cdots,\alpha_{A}])},
\end{equation}
where the summation symbol indicates the sum of different auxiliary fields $s$ and occupation configurations $\alpha_{1},\cdots,\alpha_{A}$, amplitude $A(s;\alpha_{1},\cdots,\alpha_{A})$ denotes the auxiliary field representation of Eq. \eqref{eq:smhoeq1} associated with each auxiliary field and occupation configuration, and $\mathrm{exp}(i\theta[s;\alpha_{1},\cdots,\alpha_{A}])$ is the complex phase of each pinhole amplitude. 

In practical calculations, due to the localized nature of the HO basis and its lack of translational invariance, we did not apply translations to the wavefunctions of the initial, final, or intermediate states. This may introduce the effects of CoM excitations in the calculations. However, in heavier systems, the CoM effects are suppressed because the CoM excitations are energetically prohibitive. Indeed, using the simple Hamiltonian, we found that the effect of CoM on the ground state energy of $^{22}$Si with $\tau = 0.3$ MeV$^{-1}$ is approximately $1.3\%$. In addition, each auxiliary field configuration yields a Slater determinant with a different CoM. Analysis of the CoM of the single-particle wavefunctions under each auxiliary field configuration in $^{22}$Si shows that the CoM is predominantly distributed around one lattice site near the origin. 
Moreover, the HO basis wavefunction demonstrates a good factorization property in a sufficiently large space \cite{Gloeckner:1974sst,Hagen:2009pq,Hergert:2015awm}, with the CoM part forming a Gaussian distribution. Consequently, the HO wavefunction centered at the origin is directly employed for the calculations. The truncation of the HO basis space is such that the principal quantum number $n$ is limited to $n\leq 5$ for $s$, $p$, and $d$ partial waves, and $n\leq 1$ for $f$ and $g$ partial waves.
Currently, we have primarily used occupation numbers to illustrate the shell closure. In the future, these occupation numbers will allow us to compute further physical observables associated with the shell model. 
In Tab.~\ref{tab:occ}, we show the  occupation numbers in $^{22}$Si.

\begin{table}[htb]
\centering
    \setlength{\tabcolsep}{25pt}
\begin{tabular}{lcc}
\hline\hline
\textbf{States} & \textbf{$\mathbf{n_\pi}$} & \textbf{$\mathbf{n_\nu}$} \\
\hline
$0s_{1/2}[1/2]$  & 0.935 (132) & 1.009 (123) \\\hline
$0s_{1/2}[-1/2]$ & 0.970 (135) & 0.981 (124) \\\hline
$0p_{3/2}[3/2]$  & 0.670 (151) & 0.931 (135) \\\hline
$0p_{3/2}[1/2]$  & 0.965 (164) & 0.805 (123)\\\hline
$0p_{3/2}[-1/2]$ & 0.945 (155) & 0.901 (139) \\\hline
$0p_{3/2}[-3/2]$ & 0.751 (148) & 0.911 (137) \\\hline
$0p_{1/2}[1/2]$  & 0.901 (131) & 0.877 (135) \\\hline
$0p_{1/2}[-1/2]$ & 1.049 (161) & 0.938 (147) \\\hline
$0d_{5/2}[5/2]$  & 1.018 (205)  \\\hline
$0d_{5/2}[3/2]$  & 1.056 (229)  \\\hline
$0d_{5/2}[1/2]$  & 0.980 (217) \\\hline
$0d_{5/2}[-1/2]$ & 1.106 (223)  \\\hline
$0d_{5/2}[-3/2]$ & 0.875 (209)  \\\hline
$0d_{5/2}[-5/2]$ & 1.063 (204) \\\hline
$1s_{1/2}[1/2]$  & 0.141 (173)  \\\hline
$1s_{1/2}[-1/2]$ & 0.160 (150)  \\\hline
$0d_{3/2}[3/2]$  & $-$0.127 (169)  \\\hline
$0d_{3/2}[1/2]$  & $-$0.003 (162)  \\\hline
$0d_{3/2}[-1/2]$ & 0.202 (156) \\\hline
$0d_{3/2}[-3/2]$ & 0.145 (143) \\\hline\hline

\end{tabular}
\caption{Calculated HO basis occupation numbers of protons ($\mathbf{n_\pi}$) and neutrons ($\mathbf{n_\nu}$) in $^{22}$Si based on the global chiral force $H_{\chi}$. In the States column, the HO orbitals are labeled by $nL_{j}[m]$. The values in the table represent the occupation numbers, with the numbers in parentheses indicating the statistical uncertainties.}
\label{tab:occ}
\end{table}

\subsection{Proton distances in $^{20}$Mg and $^{22}$Si}

Furthermore, we calculated the average distances of the proton distributions shown in Fig.~\ref{fig:Dis} in the main text.
The proton distributions of the inner eight protons in both nuclei exhibit significant similarity, with their average distances from the CoM differing by no more than $0.11$~fm. The largest deviation, observed in the 8th proton distribution, is likely influenced significantly by the presence of valence protons.
In $^{22}$Si, the six valence protons have average distances from the CoM of $3.22$, $3.38$, $3.51$, $3.70$, $3.97$, and $4.57$ fm, respectively. The average distances of four valence protons in $^{20}$Mg are $3.42$, $3.70$, $4.07$, and $4.74$~fm. 
Notably, the average distances of 9th and 11th protons in $^{22}$Si deviate significantly from those of the outer four valence protons in $^{20}$Mg, indicating a notable difference in spatial distribution for these particular protons.
Conversely, the 10th, 12th, and 13th protons in $^{22}$Si are close in average distance to the 9th, 10th, and 11th protons in $^{20}$Mg. This suggests that the two extra protons in $^{22}$Si are likely positioned near the regions corresponding to the 9th and 11th proton distributions.
The behavior of the outermost proton distribution between $^{22}$Si and $^{20}$Mg further highlights the distinct properties of these two nuclei.
The outermost proton distribution in $^{22}$Si has an average distance of $4.57$~fm from the CoM, which is smaller than the $4.74$~fm observed in $^{20}$Mg. This suggests that the outermost proton in $^{22}$Si is more spatially compact, even though $^{22}$Si contains more valence protons.


%
 
\end{bibunit}